# Physical Folding Codes for Proteins

Xiaoliang Ma[1], Chengyu Hou[2], Liping Shi[1], Long Li[3], Jiacheng Li[1], Lin Ye[4], Lin Yang[1,4,*] &Xiaodong He[1]

1 National Key Laboratory of Science and Technology on Advanced Composites in Special Environments, Center for Composite Materials and Structures, Harbin Institute of Technology, Harbin 150080, China
2 School of Electronics and Information Engineering, Harbin Institute of Technology, Harbin 150080, China
3 School of Electrical and Electronic Engineering, Harbin University of Science and Technology, Harbin 150001, China
4 School of Aerospace, Mechanical and Mechatronic Engineering, The University of Sydney, NSW 2006, Australia

**Abstract:** Exploring and understanding the protein-folding problem has been a long-standing challenge in molecular biology. Here, using molecular dynamics simulation, we reveal how parallel distributed adjacent planar peptide groups of unfolded proteins fold reproducibly following explicit physical folding codes in aqueous environments due to electrostatic attractions. Superfast folding of protein is found to be powered by the contribution of the formation of hydrogen bonds. Temperature-induced torsional waves propagating along unfolded proteins break the parallel distributed state of specific amino acids, inferred as the beginning of folding. Electric charge and rotational resistance differences among neighboring side-chains are used to decipher the physical folding codes by means of which precise secondary structures develop. We present a powerful method of decoding amino acid sequences to predict native structures of proteins. The method is verified by comparing the results available from experiments in the literature.

## 1. Introduction

Protein products are the basis of life on Earth and serve nearly all functions in the essential biochemistry of life science. The intrinsic biological functions of a protein are expressed and determined by its native three-dimensional (3D) structure derived from protein folding, which should be regarded as a central dogma of molecular biology [1]. Protein folding performs the phenomenon of functionalizing polymer-like polypeptides into activated proteins and bringing millions of kinds of protein into existence[2]. Protein folding can be considered the most important mechanism, principle, and motivation of biological existence, functionalization, diversity, and evolution[3]. Based on the complexity of protein folding, the protein-folding problem has been summarized in three unanswered questions[1]: (i) What is the physical folding code in the amino acid sequence that determines the particular native 3D structure rather than any other of the unfathomable number of possible conformations? (ii) What is the folding mechanism that enables proteins to fold so quickly? (iii) Is it possible to devise a computer algorithm to effectively predict a protein's native structure from its amino acid sequence? Moreover, protein folding requires aqueous environments and specific temperature ranges [4]. Here, a new algorithm, the microcanonical (NVE) ensemble relaxation engine (NVERE) is employed to reveal the mysteries of the four issues.

Protein folding is considered a spontaneous free energy minimizing process or a relaxation process that is guided mainly by the following physical forces: (i) formation of intramolecular hydrogen bonds, (ii) van der Waals interactions, (iii) electrostatic interactions, (iv) hydrophobic interactions, (v) chain entropy of protein[1]. Because molecular dynamics (MD) is capable of simultaneously describing all these physical forces and providing atomic-level resolution of protein models, MD has grown in popularity in protein-folding research since the 1980s[5,6]. Using MD to answer the three main questions of protein folding has been an enduring goal[7].

---

However, even with the MD tools, the physical folding code and the folding mechanism are still far from being understood.

More disturbingly, protein native structure predictions in MD solutions are also not encouraging. According to the laws of thermodynamics in MD, molecular systems tend towards their states of lowest free energy[1]. But as yet, predictions of protein native structures tend to require a significant amount of computational resources and have succeeded only for a limited set of small protein folds[5,8,9]. Existing algorithms have been unable to simulate a protein folding following an explicit folding code, often resulting in misfolding during the search for folding pathways[10]. The strategy of repeatedly misfolding and resampling adopted by these algorithms may well be inconsistent with the way nature works in protein folding. Thus, existing algorithms should be interpreted as protein-folding searching strategies rather than protein-folding simulations, so that protein native structure prediction still remains extremely challenging.

Experimental evidence has shown that a protein appears first to develop some local structures (such as helices and turns) in the main chain, followed by growth into more global structures (such as α-helices and β-sheets) [11]. Laser temperature-jump studies have experimentally determined that α-helices and β-sheets form very quickly at microsecond timescales[12,13]. These findings all indicate that there may be an explicit folding code dominating the folding process or pathway of a given protein.

Protein folding often does not occur in isolation but in an aqueous solution, and proteins denature (i.e., lose their native structures) in most non-aqueous solvent [14,15]. The answers to the three main questions may lie in a deeper understanding of the physical and chemical difference between neighboring side-chains along the amino acid sequences of proteins in aqueous environments. The way that water molecules interact with unfolded proteins explains one of the great mysteries of protein folding. Water molecules are characterized by small size, negatively and positively charged at O and H respectively. This means that small water molecules can squeeze into the gap between neighboring side-chains and hydrogen bond with the H-N group and C=O group of each hydrophilic side-chain. Thus, water molecules should be able to saturate the hydrogen bond formations of hydrophilic side-chains. Surrounding water molecules most likely shield a hydrophilic side-chain from hydrogen bonding with the main chain and the other hydrophilic side-chains, thus preventing the side-chains from interfering with the formation of the secondary structure. Experiments[16,17] have also shown that secondary structures (such as α-helices and β-sheets)  are mainly stabilized by hydrogen bonds between the N-H groups and C=O groups of the main chain. Thus, this problem lies in our lack of understanding of the properties of unfolded proteins in aqueous environments. Because of this paucity of information, some thermodynamically important states of unfolded proteins may be overlooked [18].

In aqueous environments, conformation of an unfolded protein may first enter an unknown metastable state to facilitate activation of the unknown physical folding code (Fig.1). The peptide group (CO–NH) has a rigid planar structure. Peptide (C-N) bonds (shaded yellow in Fig. 1a), that make up one-third of all backbone bonds, have some double-bond characteristics due to resonance and are not free to rotate. Note that each carbonyl oxygen atom in a planar peptide group tends to bond with an amide hydrogen atom in an adjacent planar peptide group due to the electrostatic attractions between them, causing a tendency for the C=O group and N-H group to be parallel to each other (see Fig. 1a). Thus, the other bonds in the backbone may also be rotationally hindered. This indicates that an unfolded protein in an aqueous environment may be unable to fold spontaneously unless thermal motion breaks the electrostatic attraction between the C=O groups and the H-N groups of the main chain. Aqueous environments and appropriate temperatures may work cooperatively to facilitate activation of the explicit physical folding codes for proteins.

Here, we use an NVERE to investigate the conformational development and mechanical behaviors of unfolded proteins in aqueous environments. We show that relaxed, unfolded proteins in aqueous environments are characterized by adjacent planar peptide groups parallel to each other as the result of electrostatic attraction between the C=O group and H-N group in every pair of adjacent planar peptide groups. Temperature-induced torsional waves propagating along the main chain are found to be an underlying mechanism responsible for the development of secondary structures. Torsional resistance in aqueous environments and electric charge differences between neighboring side-chains are inferred as the keys for cracking the physical folding code. These characteristics enable an appropriate temperature environment to break the parallel distributed state of some specific amino acids in torsion failure, thereby activating the physical folding code and the initiation of protein folding. We perform parametric analyses to identify several types of amino-acid sequence codes (i.e., physical folding codes) that lead to the initiation or cessation of the formation of secondary structures.

On that basis, we develop a method for decoding amino acid sequences to predict the folding processes of proteins, verifying the method by comparing the results with those from experiments[16,17]. We then use the canonical ensemble (NVT) to simulate the rapid formation of secondary structures. The underlying mechanism responsible for the superfast folding of protein is found to be that each hydrogen bond formation can speed the formation of the next hydrogen bond.

## 2. Results

To demonstrate the shielding effect of water molecules on hydrophilic side-chains, we first try to minimize the potential energy of several native protein structures in isolation, using the CHARMM[19] force field and the NVERE (see Supplementary S1). The NVERE relaxation features the optimization of potential energy through long MD trajectories and large deformation[20]. It is surprising that the potential energy of these native protein structures is further minimized, making the structures more compact through the formation of more intramolecular hydrogen bonds (see Supplementary Fig. S1, Movie S1, and Movie S2). This result indicates that the structures of these proteins are not sterically permitted in nature, and have not fully exploited their intramolecular hydrogen-bonding capacity[21]. These native structures are experimentally determined in aqueous solution and should be the states of lowest free energy in water rather than in isolation. This means that aqueous environments are capable of shielding hydrophilic side-chains from hydrogen bonding with other hydrophilic side-chains and the main chains.

To simulate the shielding effect of aqueous environments, a procedure in nature eliminates the atomic charges of side-chains. A molecule model of an unfolded protein (1B64) is built with side-chains and planar peptide groups randomly distributed, as shown in Fig. 2a. We use the NVERE to minimize the potential energy of the unfolded protein and simulate the conformational development of the unfolded protein (see Supplementary S2). The relaxation results show that the adjacent planar peptide groups gradually become parallel to each other as the result of electrostatic attraction between the C=O group and the H-N group in the main chain (Fig. 2a). Parallel distribution of adjacent planar peptide groups also causes adjacent side-chains to distribute on opposite sides of the main chain and each side-chain is parallel to every other side-chain, making the whole relaxed unfolded protein look like a 'centipede kite' (see Figs. 1c, 2a and Supplementary Movie S3). It is worth noting that this centipede-kite-like state of the unfolded protein is very similar to the molecular configuration of β-sheets (except for the turns) in native protein structures, as illustrated in Fig. 1b and 1c. This resemblance indicates that the centipede-kite-like state of unfolded protein may be a large fraction in proteins' native structures.

To test the stability of the centipede-kite-like state of unfolded protein, we apply torsion loading on one end of the centipede-kite-like structure. Deformation of the unfolded protein under the common loads of torsion is obtained in an NVE ensemble. It is surprising that the unfolded protein is capable of transmitting torsion moment along itself without breaking the centipede-kite-like state (except for the proline) as shown in Fig. 2b, 2c and Supplementary Movie S4. To reveal the deformation behaviors of the centipede-kite-like structure under thermal loading, we put the structure in an NVT (conserving of substance, volume, and temperature) ensemble at 35°C. The simulation results show that the thermal environment arouses mainly torsional vibrations of side-chains and torsional waves propagating along the main chain, as shown in Supplementary Movie S5. A higher temperature arouses torsional waves with more volatility. The results indicate that an unfolded protein in water cannot spontaneously fold unless a thermal environment breaks the electrostatic attractions between the C=O groups and H-N groups with some specific amino acids.

The answer to the physical folding code must lie in a deeper understanding of the reason for thermal environment breaking the electrostatic attractions between the C=O groups and H-N groups with these specific amino acids in unfolded protein during folding. Rotational resistance and electric charge differences among neighboring side-chains should be considered as the keys. Obviously, the torsional resistance of a long hydrophilic side-chain (such as arginine) played out in water should be much greater than that of a short hydrophobic side-chain (such as glycine) in a waterless condition, as shown in Fig.1d. The rotational inertia of a long side-chain about the axis of the backbone is much greater than that of a short amino acid. Hydrogen bonding between a hydrophilic side-chain and surrounding water molecules can significantly increase the rotational resistance of the side-chain in aqueous environments. Thus, the rotational resistance of a side-chain can be evaluated by its rotational inertia and the hydropathic index[22] of the amino acid, as shown in Supplementary Table S1, in which 20 kinds of amino acid are divided into six categories according to their rotational resistance: strong resistance ($R_s$), medium resistance ($R_m$), weak resistance ($R_w$), very weak resistance ($R_{vw}$), glycine (G) and proline (P).

Moreover, nine amino acids with electrically charged side-chains are subdivided into three other categories, negative ($N_e$), negative weak ($N_{ew}$), and positive ($P_o$), as shown in Supplementary Table S2. Electrical repulsion between neighboring electrically charged side-chains might be able to cause the side-chains to rotate synchronously when rotational waves pass through. For example, when two $P_o$ amino acids (**$P_o$-$P_o$**) are in sequence, the two positive side-chains should alternate opposition and repulsion with each other. When a rotational wave passes through the first $P_o$, the side-chain of the second $P_o$ also begins to rotate due to the repulsion between the side-chains, thereby grouping the two amino acids together and doubling the rotational resistance (see Supplementary Fig. S2a). Thus, we can conceive the **$P_o$-$P_o$** as a group characterized by ultrastrong rotational resistance. We identify four kinds of double amino acid groups that can provide ultrastrong rotational resistance and label them **$R_u$**, as illustrated in Supplementary Table S3.

When two amino acids with a very large difference in rotational resistance are coded together in chain, the two neighboring amino acids may be able to block torsional waves passing through and result in torsion failure of the parallel distributed state, namely folding. For example, a sequence segment of **$R_s$/$R_u$-G** should be a typical weak point of an unfolded protein that could result in breakage of the parallel distributed state with torsion failure caused by torsional waves. The rotational resistance of G is significantly lower than that of the **$R_s$/$R_u$** amino acid. When torsional waves pass through the **$R_s$/$R_u$-G** segment, the amplitude of the torsional vibration aroused in **G** would obviously be greater than that of the adjacent **$R_s$/$R_u$** amino acid, potentially resulting in breaking the electrostatic attraction between the C=O group and the H-N group of G (see Figs. 1d). The folding should be initiated at the **G**. Therefore, **$R_s$/$R_u$-G** in a sequence could be used as a simple physical folding code for predicting the formation of a turn in a secondary structure. If that code exists widely in the turns including the β sheets turns in protein structures, we may be able to prove that the code exists and works universally. We identify 821 codes of **$R_s/R_u$-G** in 124 small proteins using the Protein Data Bank archive (PDB) and find that about 96% of these codes are present in the turns including the β sheets turns (see Supplementary S10). Moreover, we find that about 60% of β sheets are characterized by the presence of **$R_s/R_u$-G** at the turns in protein native structures, indicating that this code exists extensively.

It is also worth noting that proline (symbol **Pro** or **P**) does not contain the N-H group (Fig. 1d and 2c). The torsional strength of proline is negligible compared to that of other amino acids due to the lack of electrostatic attraction. **P** should be considered another simple physical folding code for predicting turns in secondary structures of protein. Using the PDB, we identify 387 codes of **P** present in the 124 proteins and find that about 99% of **P** codes in these proteins can cause a turn to be generated, as shown in Fig. 2c and Supplementary S10. Moreover, about 14% of the turns in β sheets are characterized by **P**, indicating that the **P** code also exists extensively.

Moreover, temperature generated rotational waves may be able to trigger some electrically attractive motions in neighboring electrically charged side-chains, thereby breaking the parallel distributed states of these amino acids and initiating folding. When a $P_o$ amino acid and an $N_e$ amino acid are coded together in sequence ($P_o$-$N_e$), the two side-chains attract each other, which may result in their rotation (see Supplementary Fig. S2b). A $P_o$ amino acid adjacent to an aspartic (D) ($P_o$-D) should be a typical physical folding code that may be able to drive attractive motion between them. We identify 130 codes of $P_o$-D in the 124 proteins and find that about 97% of the $P_o$-D codes in these proteins are present in turns, see Supplementary S10. This finding indicates that the $P_o$-D physical folding code can result in the generation of turns.

Attractive motions of charged side-chains around the main chain may contribute to the α helix formation. We identify three electrostatically driven rotation codes (**EDRC**) that can result in remarkably attractive motions of side-chains around the backbone, as illustrated in Supplementary Table S4. The formation of a helix can be conceived as a spiral process and a contractive process of a segment of polypeptide. We identify four patterns of long-range electrostatic attraction (**LEA**), as illustrated in Supplementary Table S5. The LEA most likely contributes to the formation of the first helix through contracting a segment of five amino acids that have lost their parallel distributed states. An EDRC, one or two $R_{vw}$, and an **$R_u/R_s$** (EDRC-$R_{vw}$- **$R_u/R_s$**) coded together with a strong LEA between the EDRC and **$R_u/R_s$** may be a typical code of an α helix formation. Electrostatically driven rotational waves generated by the EDRC can be blocked by the **$R_s/R_u$** after they pass through the $R_{vw}$. This phenomenon could cause all five amino acids in the EDRC- $R_{vw}$- **$R_s/R_u$** to lose their parallel distributed states. We identify 126 EDRC-$R_{vw}$-**$R_s/R_u$** codes in 147 small proteins and find that 90% of the EDRC-$R_{vw}$- **$R_s/R_u$** codes in these proteins are present in α helices, see Supplementary S7 and S10. Moreover, about 29% of the helices are characterized by the EDRC-$R_{vw}$- **$R_s/R_u$**, indicating that the EDRC-$R_{vw}$- **$R_s/R_u$** code also exists extensively.

Through analyzing protein structures in the PDB, we identify at least three types of physical folding code that can result in the formation of an α helix. The three codes and their requirements are given in Supplementary Table S7. We also identify six types of physical folding code that can results in formation of a turn including the β sheet turns. The six codes and their requirements are given in Supplementary Table S6.

We compare our predictions of 416 proteins with those from experiments in the PDB. The results show that these codes have a high success rate in predictions of α helices (about 80 %) and turns including β sheet turns (about 95%) in these proteins. The results of predictions of the two proteins are illustrated in Fig. 3; other results are available in Supplementary S10. Our preliminary results show that the physical folding codes deduced from torsional resistances in aqueous environments and the electric charge performances of side-chains are powerful in predicting secondary structures.

Both experimental evidence[11] and our preliminary results indicate that protein folding initiates from the development of a local structure (such as a helix or a turn). We use a canonical ensemble (NVT) to simulate the folding processes of a β sheet and an α helix based on an unfolded conformation with a turn and a short helix, respectively (see Fig. 4). The simulation results show that the spontaneous protein-folding process and the development of the unfolded protein are guided mainly by the continuous formation of hydrogen bonds (see Supplementary Movies S6 and S7). Each hydrogen bond formation can bring another pair of a C=O group and an N-H group close enough to electrostatically attract each other, further facilitating the formation of the next hydrogen bond. We show that the formation of intramolecular hydrogen bonds increases the folding power during formation of the β-sheet and the α helix. The simulation results also indicate that the folding power accumulated during the formation of secondary structures pulls these secondary structures closer together, leading to the formation of the tertiary level structure, resulting in the superfast folding of the whole protein.

## 3. Conclusion

In conclusion, preliminary answers to the three main questions of protein folding are given based on the physical analysis. The parallel distributed state of adjacent planar peptide groups of unfolded protein must be an indispensable intermediate that enables a polypeptide to fold accurately in aqueous environments and a large fraction to remain in β sheets of native structures. The physical folding code can be deciphered through evaluating torsional resistance and electrical charge differences of neighboring side-chains. The validity of these codes has been confirmed by comparing the simulated results with those from experiments[16,17]. Protein folding is dependent on the shielding effect of water molecules as the way to pave a native folding pathway. Temperature-induced torsional waves propagating along the centipede-kite-like structure of unfolded protein should be considered as the key activating the physical folding code. On the basis of our results, inserting or removing a key amino acid (such as P or G) in or from the sequence could completely change the secondary structure and influence the configuration of the tertiary structure of a protein. The existence, functionalization, diversity, and evolution of millions of kinds of protein on Earth must have been derived from the delicate coding strategies that fully utilize the differences in torsional resistance and in the electrical charges of side-chains. Our method will be useful for predicting and evaluating structures and biological functions of protein based on amino acid sequences and may contribute to improving the success rate of new drug designs. This research provides a new direction for exploring the evolution of proteins on Earth, promising innovations in both biochemistry and molecular biology.

## Methods

The NVERE is employed to simulate the conformational development of unfolded proteins through eliminating the atomic charge of side-chains by virtue of the shielding effect of aqueous environments. In general, the unfolded states adopted by polypeptide chains are believed to be physically stable configurations in which the free energy/potential energy is minimized. Structure relaxation via a longer trajectory should be particularly appealing for relaxing the unfolded protein working with soft long chains. A variety of MD methods, such as conjugate gradient, Newton-Raphson, QM, and FIRE, have been established to solve the relaxation task that systematically removes potential energy from a system [23-25]. However, for the protein-folding problem, these MD methods often do not permit longer MD trajectories that could identify a global equilibrium through large deformation, namely that energy minimization generally leads to a state that is composed of many local minima [26]. The NVERE is a simple method of utilizing the NVE ensemble to obtain static equilibriums for molecular systems at very fast rates[20,27]. In particular, the method permits longer MD trajectories towards lower energy basins and finding more stable equilibriums. In this study, all MD simulations were conducted with the same time step (2 fs per time step). The method is well suited to problems quite distinct from the static equilibrium. The NVERE is capable of finding more stable equilibrium configurations than common optimization algorithms. The simulations were executed using the LAMMPS (large-scale atomic/molecular massively parallel simulator) MD simulator with the CHARMM force field[28,29]. The formation of intramolecular hydrogen bonds was monitored using PyMOL, enabling us to depict atomic-level resolution of hydrogen bond distribution throughout the polypeptide.

**Figures:**

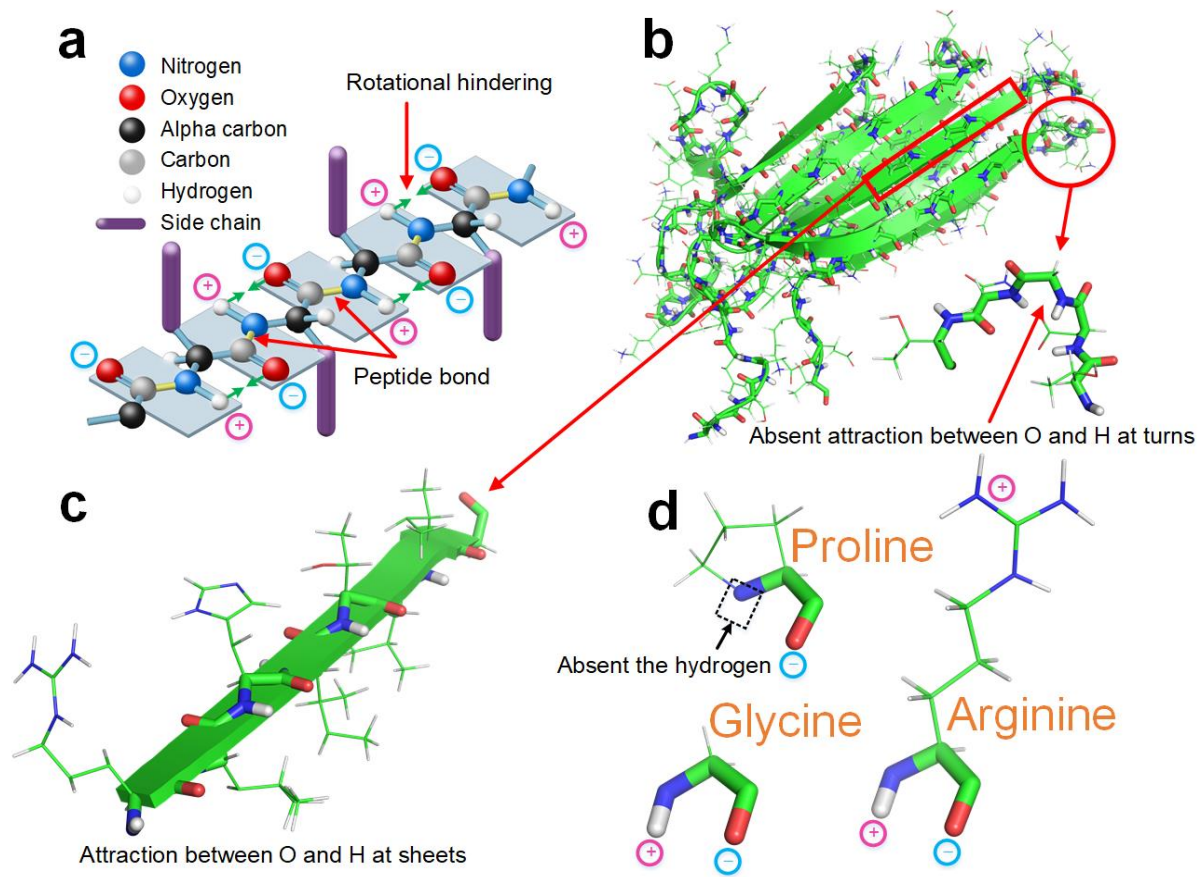

Figure 1. Parallel distributed state of the planar peptide groups of an unfolded protein. **a,** All three bond types in a polypeptide chain rotationally hindered. The peptide bonds in the planar peptide groups are shaded yellow. Each peptide bond has some double-bond character due to resonance and cannot rotate. The carbonyl oxygen has a partial negative charge and the amide nitrogen has a partial positive charge. The two other bond types in the backbone may also be rotationally hindered, depending on electric attraction between the carbonyl oxygen and amide nitrogen. **b**, Protein 2LRG. **c,** Parallel distributed state of planar peptide group in a β sheet of the protein 2LRG. **d,** Proline, glycine, and arginine.

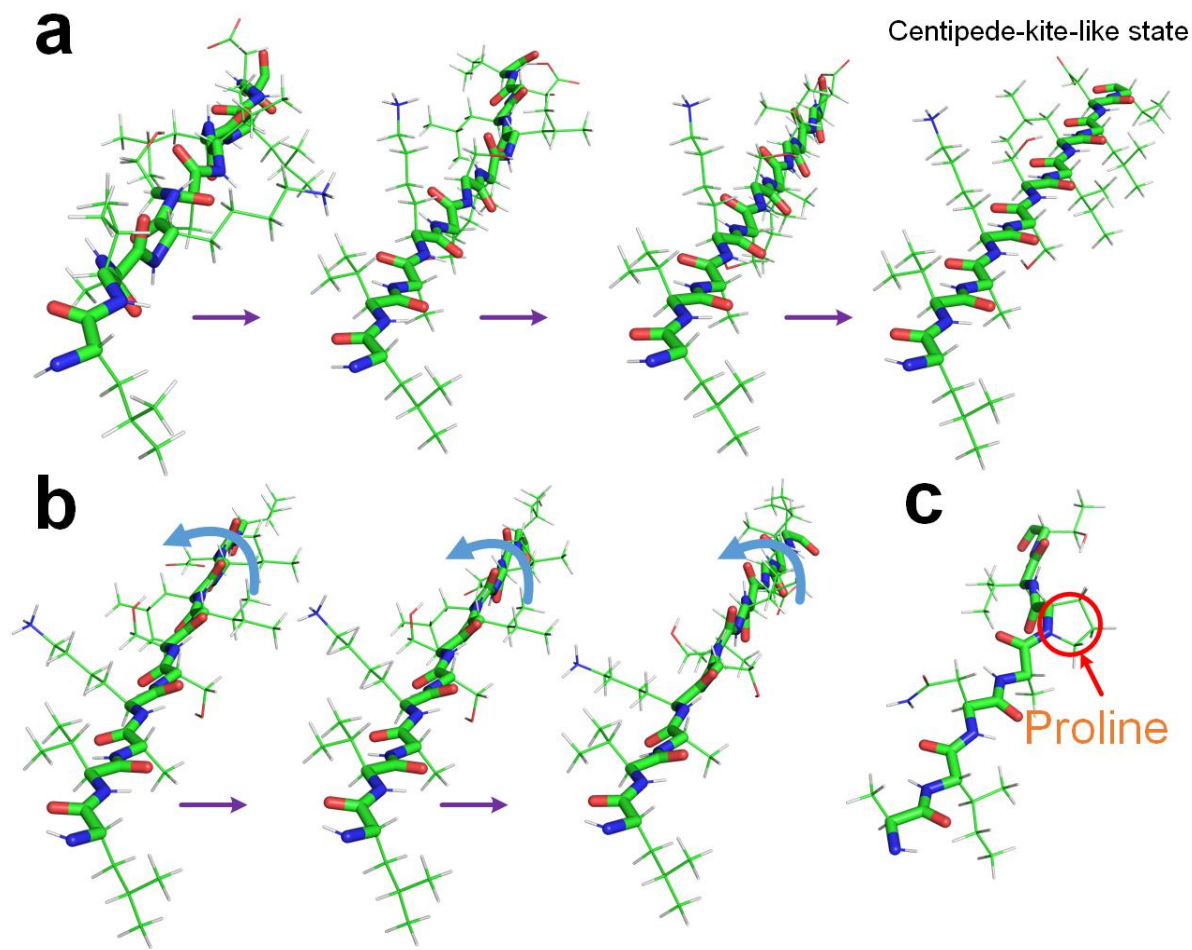

Figure 2. Simulated optimization of potential energy of an unfolded protein. **a,** Conformational development of the unfolded protein (1B64) in relaxation. **b,** Deformation of the centipede-kite-like unfolded protein (1B64) under torsion. **c,** Lack of electrostatic attraction bonding between the N-H group and C-O group and parallel distributed pattern of proline in protein 1DOL.

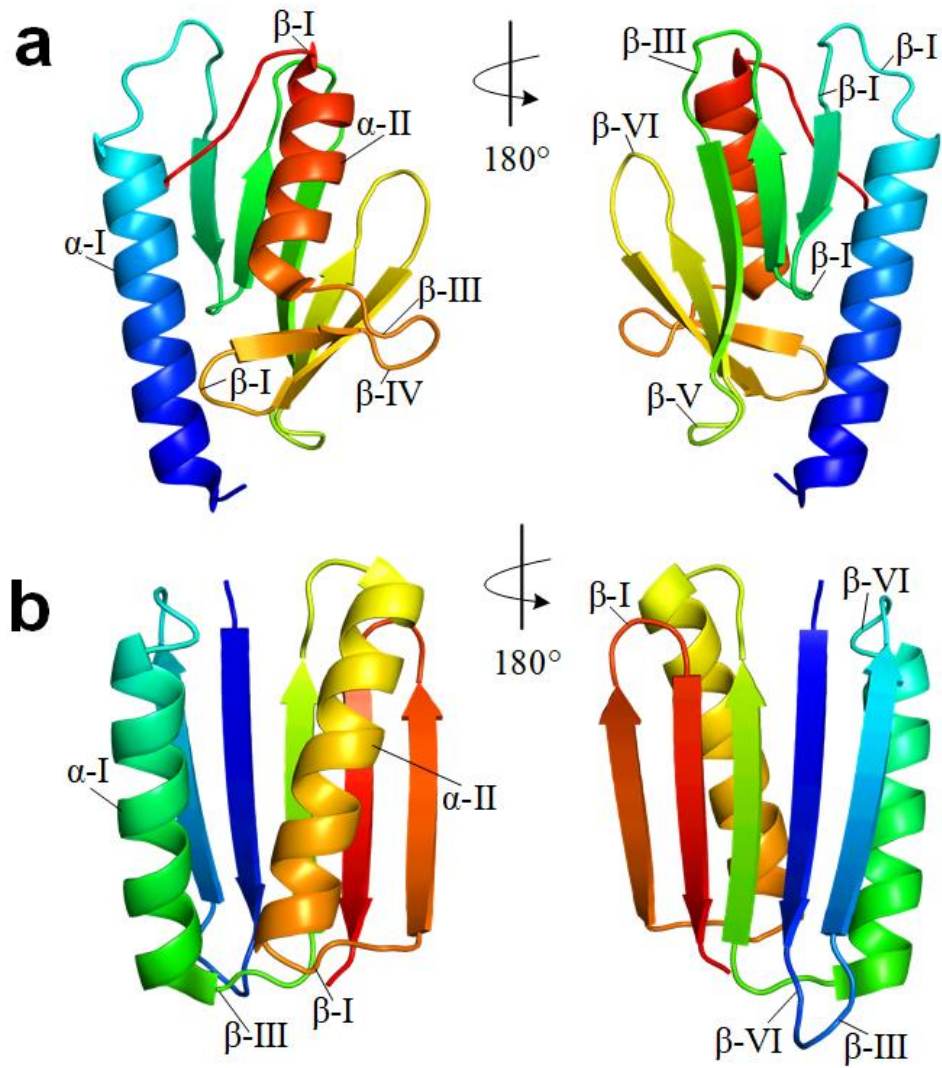

Figure 3. Results of secondary structure predictions of two proteins. **a,** 1EW4. **b,** 1QYS. The α helix codes are denoted by α and the turn codes are denoted by β.

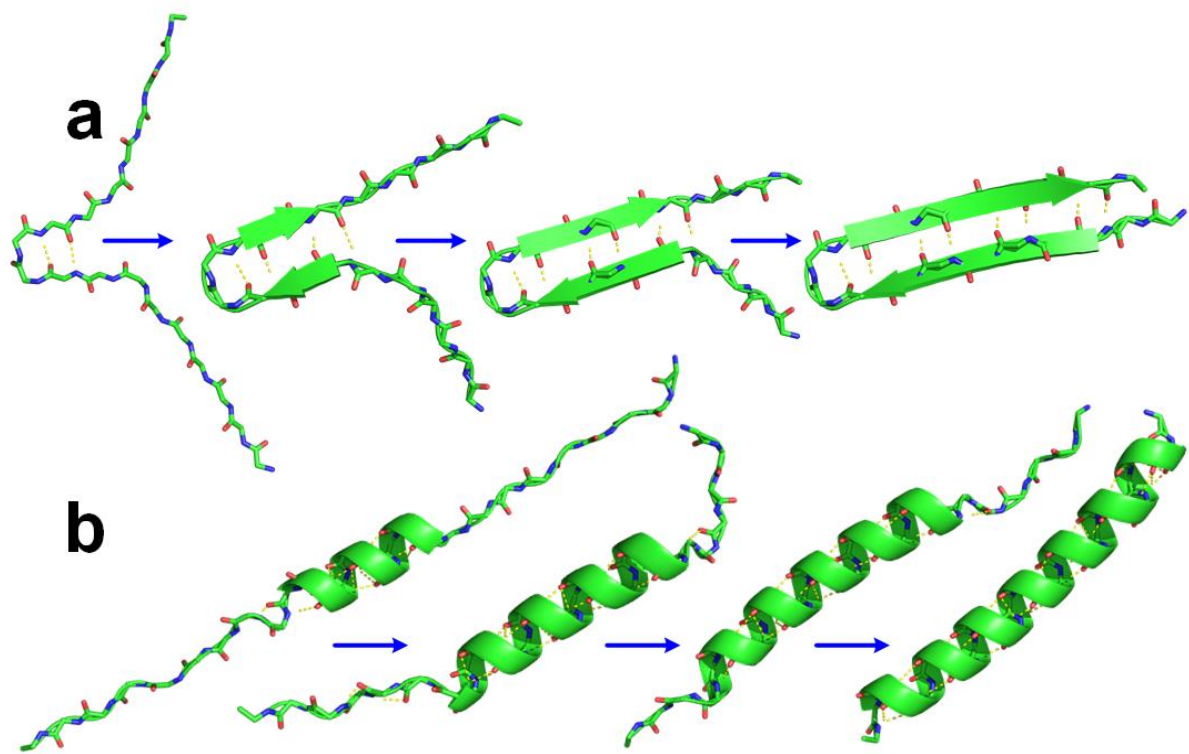

Figure 4. Rapid development of **a**, β-sheet and **b**, α helix

**Supplementary Materials**
**Note S1. Shielding effect of water molecules on hydrophilic side chains**
To demonstrate the shielding effect of water molecules on hydrophilic side chains, we first tried to minimize the potential energy of ten native protein structures (1WHX, 2H7A, 2CQD, 2DM8, 1WJ3, 2CSJ, 2LJK, 2EO9, 2DAF and 2NBB) in isolation, using the CHARMM force field and the NVERE. NVERE relaxation incorporates features optimizing potential energy through long MD trajectories and large deformation. It was surprising that the potential energies of the native structures of proteins were further minimized, making the structures more compact through the formation of more intramolecular hydrogen bonds (as illustrated in Supplementary Fig. S1, Movie S1 and Movie S2). This indicates that the native structures of these proteins are not sterically allowed and most fully exploit the intramolecular hydrogen-bonding capacity of these structures[1]. The roles played by surrounding water molecules in protein folding must be closely related to the distinctive physical and chemical properties of water molecules. These native structures are experimentally determined in aqueous solution and should be the states of lowest free energy in water rather than in isolation. The simulation results show that aqueous environments are capable of shielding hydrophilic side-chains from hydrogen bonding with other hydrophilic side-chains and the main chains (Fig. S1a and S1b). The potential energy versus time curves in the NVERE algorithms are also plotted in Fig. S1a and S1b.

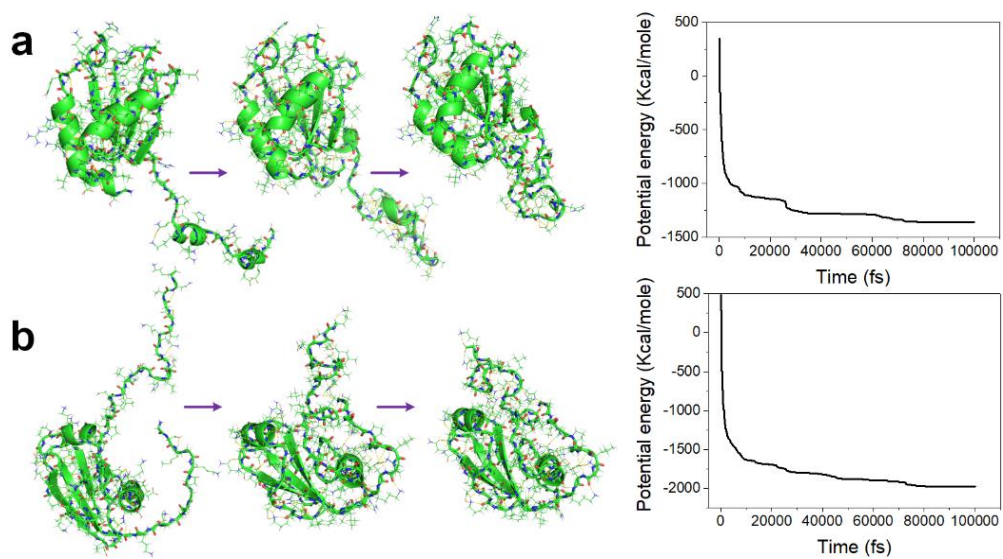

Figure. S1. **a**, Molecular configuration variations of protein 1WHX during relaxation using NVERE and optimization of the potential energy of 1WHX during relaxation. **b,** Molecular configuration variations of protein 2H7A during relaxation using NVERE and optimization of potential energy of 2H7A during relaxation.

**Note S2. Relaxation of unfolded protein in water**
At the onset of protein-folding, the folding process would be not guided by the formation of intermolecular hydrogen bonds, due to the shielding effect of water molecules. A reasonable protein-folding simulation must at least be able to avoid the development of hydrogen bonding interaction of side chains at beginning. Unfortunately, it is difficult to directly simulate the structural development of an unfolded protein in aqueous environment because water molecular movement is instantaneous and the formation and breakage of such hydrogen bonds among the unfolded protein and surrounding molecules causes potential fluctuations within the system. To simulate the shielding effect of aqueous environments, a simple method is to eliminate the atomic charge of side chains.

Several complex force fields, such as Chemistry at Harvard Macromolecular Mechanics (CHARMM)[2], Assisted Model Building with Energy Refinement (AMBER)[3], DREIDING[4], and GROningen Molecular Simulation (GROMOS)[5] have been developed for protein molecular structure, making it possible to determine at the atomic level which mechanisms are guiding the protein-folding process or which folding intermediates (i.e., partially structured states) along the folding pathway could give insight into the physical folding code[6]. In our current simulation using the CHARMM force field, a molecule model of an unfolded protein (1B64) was built with side-chains and planar peptide groups randomly distributed, as shown in Fig. 2A and Movie S3. The relaxation results show that the adjacent planar peptide groups gradually became parallel to each other as the result of electrostatic attraction between the C=O group and the H-N group in the main chain (see Movie.S3).

**Note S3. Rotational resistance performance of different amino acids**

Obviously, the torsional resistance of a long hydrophilic side-chain (such as Arginine) played out in water should be much greater than that of a short hydrophobic side-chain (such as Glycine), as shown in Fig. 1d. Because the rotational inertia of a long side chain about the axis of the backbone is much greater than that of a short amino acid. The hydrogen bonding between a hydrophilic-side chain and surrounding water molecules can significantly increase the rotational resistance of the side-chain in aqueous environments. Thus, the rotational resistance performance of a side-chain can be evaluated by its rotational inertia and the hydropathic index of the amino acid, as shown in Table S1, in which 20 amino acids are divided into six categories according to their rotational resistance, classed as strong resistance ($R_s$), medium resistance ($R_m$), weak resistance ($R_w$), very weak resistance ($R_{vw}$), Glycine (G), and Proline (P). It worth noting that all amino acids in the $R_{vw}$ category are characterized by a lower moment of inertia and are hydrophobic.

| Amino acid | Moment of inertia/Å$^2$ | Hydropathy index[22] | Group |
|---|---|---|---|
| R | 3278.58 | -2.5 | $R_s$ |
| K | 1613.13 | -1.5 | |
| Q | 1308.01 | -0.85 | |
| E | 1277.22 | -0.74 | |
| H | 1387.18 | -0.40 | |
| N | 591.66 | -0.78 | |
| D | 599.73 | -0.90 | |
| T | 364.59 | -0.05 | $R_m$ |
| S | 198.17 | -0.18 | |
| Y | 1944.75 | 0.26 | $R_w$ |
| W | 2779.20 | 0.81 | |
| C | 406.95 | 0.29 | |
| M | 1379.63 | 0.64 | |
| F | 1308.47 | 1.2 | |
| L | 712.65 | 1.1 | $R_{vw}$ |
| A | 76.53 | 0.62 | |
| V | 360.45 | 1.1 | |
| I | 641.86 | 1.4 | |
| G | 2.93 | 0.48 | — |
| P | 161.06 | 0.12 | — |

Table S1. Rotational resistance of side-chains of amino acids

**Note S4. Electrically charged side-chains**

Twelve amino acids with electrically charged side-chains are subdivided into three other categories, negative ($N_e$), negative weak ($N_{ew}$), and positive ($P_o$), as shown in Table S2.

| Amino acid | E | D | N | T | Q | S | R | K | H |
|---|---|---|---|---|---|---|---|---|---|
| Isoelectric point | 3.15 | 2.85 | 5.41 | 5.60 | 5.65 | 5.68 | 10.76 | 9.60 | 7.60 |
| Group | $N_e$ | | $N_{ew}$ (and Hydrophilic) | | | | $P_o$(Positive) | | |
| | $N_{ea}$(Negative all) | | | | | | | | |

Table S2. Electrical charge of side-chains of amino acids.

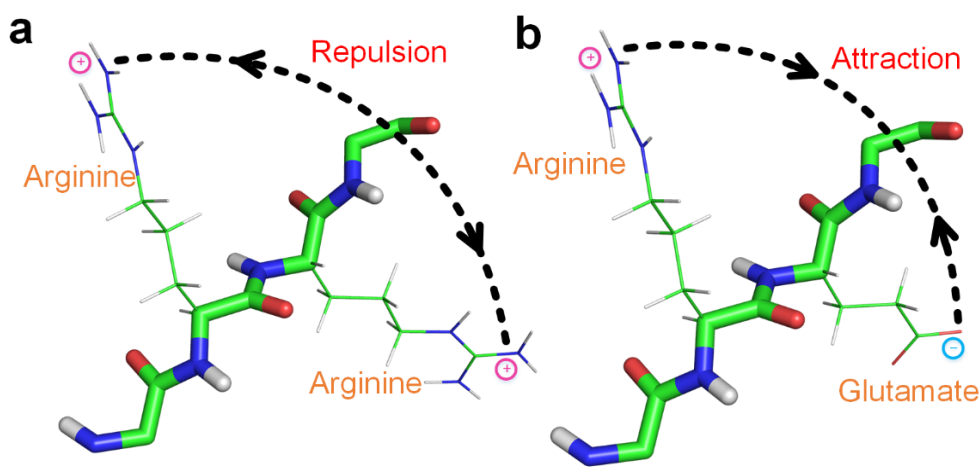

Figure. S2. Illustration of electrical repulsion and attraction between neighbored electrically charged side chains. **a**, Repulsion. **b**, Attraction.

**Note S5. Amino acid groups with ultrastrong rotational resistance**

Electrical repulsion between neighboring electrically charged side-chains can cause the side-chains to rotate synchronously when rotational waves pass through them. For example, when two $P_o$ amino acids ($P_o$-$P_o$) are in sequence, the two positive side-chains should distribute opposition and repulsion with each other. When a rotational wave passes through the first $P_o$, the side-chain of the second $P_o$ also begins to rotate due to the repulsion between the side-chains, causing make the two amino acids to group together and double the rotational resistance, as shown in Fig. S2a. Thus, we can conceive the **$P_o$-$P_o$** as a group characterized by ultrastrong rotational resistance. We identified four kinds of double amino acid groups that could provide ultrastrong rotational resistance and labeled them **$R_u$**, as illustrated in Table 3.

|  | Group |
|---|---|
| **$R_u$** 1 | $N_e$-$N_e$ |
| **$R_u$** 2 | $P_o$-$P_o$ |
| **$R_u$** 3 | $N_e$-$N_{ew}$ |
| **$R_u$** 4 | $N_{ew}$-$N_{ew}$ |

Table S3. Amino acid groups providing ultrastrong rotational resistance

**Note S6. Electrostatically driven rotation codes (EDRC)**

We identified three **EDRC**s that could result in remarkably attractive motions of side chains around the backbone, as illustrated in Table S4.

| Electrostaticlly driven rotation codes | |
|---|---|
| EDRC 1 | $\vec{P_o} - \overleftarrow{N_{ew}}$ |
| EDRC 2 | $\vec{P_o} - \overleftarrow{N_e}$ |
| EDRC 3 | $\vec{P_o} - \overleftarrow{N_{ew}}|N_e - \overleftarrow{N_{ew}}|N_e$ |

Table S4. Electrostatically driven rotation codes

**Note S7. Long-range electrostatic attraction (LEA)**

The formation of a helix can be conceived as a spiral and contractive process of a segment of polypeptide. Thus we speculate that there is some LEA between charged side-chains that can contribute to the α helix formation. We identified four patterns of LEA as illustrated in Table S5. LEA most likely contributes to the first helix formation through contracting a segment of amino acid after the parallel distributed states are lost.

| | Long-range electrostatic attraction | |
|---|---|---|
| 1 | LEAS | $\vec{P_o} - U_n^1 - U_n^2 - \overleftarrow{N_e}$ |
| 2 | (LEA strong) | $\vec{P_o} - U_n^1 - U_n^2 - U_n^3 - \overleftarrow{N_e}$ |
| 3 | LEAW | $\vec{P_o} - U_n^1 - U_n^2 - \overleftarrow{N_{ew}}$ |
| 4 | (LEA weak) | $\vec{P_o} - U_n^1 - U_n^2 - U_n^3 - \overleftarrow{N_{ew}}$ |

Table S5 Long-range electrostatic attraction

**Note S8. Codes for turn prediction**

Through analyzing protein structures from the PDB, we identified that there are at least exist six amino acid code sequences that could result in the formation of turns. The six kinds of codes and the corresponding requirements are shown in Table S6. We also found that the codes I and V are the most common code for the formation of turns. Short β sheets seldom appear at both ends of a protein's native structure because both ends of an unfolded protein are more rotationally free. These findings indicate that parallel distributed states of these amino acids at both ends of an unfolded protein are difficult to break by thermal motion. So these codes in Table S6 do not apply to the ends of amino acid sequences of proteins. The folding mechanism for code I is the G-induced large difference in rotational resistance between neighboring side-chains along amino acid sequences of proteins in aqueous environments. The folding mechanism for code II is the $R_{vw}$-induced large rotational resistance difference between neighboring side-chains along the amino acid sequences of proteins in aqueous environments. The folding mechanism for code III is the $P_o$-$N_{ew}$-induced electrically attractive motions of neighboring electrically charged side-chains. The folding mechanism for code IV is the $P_o$-$N_e$-induced electrically attractive motions of neighboring electrically charged side-chains. The folding mechanism for code V is the torsional strength of P, which is negligible compared to that of other amino acids due to the lack of electrostatic attraction. The folding mechanism for code VI is electrically repulsive motions of neighboring electrically charged side-chains on the same side as the main chain.

| | | Codes | Requirements |
|---|---|---|---|
| I | 1 | $R_u\|R_s - G - R_s\|R_u\|R_m$ | Except $R - G - C$ |
| | 2 | $R_u\|R_s\|R_m\|G\|Y - R_{vw}\|R_w - G - R_u\|R_s\|R_m$ | $R_s\|R_m \notin EDRC1\|2$ |
| | 3 | $G - G$ | |
| | 4 | $R_u\|R_s - R_{vw} - G - R_{vw} - R_u\|R_s\|R_m$ | |
| | 5 | $R_u\|R_s\|G - R_{vw} - R_{vw}\|R_w - G - R_u\|R_s\|R_m - U_n^1$ | $U_n^1 \neq G$, Except $EDRC1 - R_{vw} - R_{vw} - G - EDRC1\|2$ |
| | 6 | $R_s\|R_m - G - R_m\|R_w - R_m\|R_s$ | |
| | 7 | $U_n^1 - EDRC1\|2 - G - R_w\|R_m - U_n^2$ | |
| | 8 | $G - R_{vw} - R_w\|R_{vw}\|R_m - G$ | |
| | 9 | $R_u\|R_s\|G - R_{vw} - R_{vw} - R_{vw}\|R_w - G - R_u\|R_s\|R_m$ | |
| | 10 | $N_e\|N_{ew} - R - R_{vw} - G$ | |
| II | 1 | $U_n^1 - R_u^1 - R_{vw} - R_u^2 - U_n^2$ | $R_u^1 \neq R_u3\|R_u4, R_u^2 \neq R_u4, U_n^1 \neq P, U_n^2 \neq P$ |
| | 2 | $U_n^1 - R_u^1 - R_{vw} - R_{vw} - R_u^2 - U_n^2$ | $R_u^1 = R_u2, R_u^2 \neq R_u3\|R_u4, U_n^1 \neq P, U_n^2 \neq P$ |
| | 3 | $N_e\|N_{ew} - R - R_{vw} - P_o - N_{ew}$ | $P_o \neq H$ |
| | 4 | $N_e\|N_{ew} - R - R_{vw} - R_u^1$ | $R_u^1 \neq R_u4$ |
| | 5 | $R - N_e\|N_{ew} - R_{vw} - R_u^1$ | $R_u^1 \neq R_u4$ |
| | 6 | $EDRC2 - R_{vw} - EDRC2$ | |
| III | 1 | $P_o - N$ | $P_o \neq H$ |
| | 2 | $G - N_{ew} - P_o$ | |
| | 3 | $U_n^1 - N_{ew} - P_o - G$ | $U_n^1 \neq P_o$ |
| | 4 | $U_n^1 - N_e - U_n^2 - P_o - U_n^2$ | $U_n^1 \neq P_o, U_n^2 = S\|T\|Q, U_n^2 \neq N_e$ |
| IV | 1 | $P_o - D$ | |
| | 2 | $U_n^1 - P_o - E - U_n^2$ | $P_o \neq H, U_n^2 \neq N_e, U_n^1 \neq R_{vw}, U_n^2 \neq R_{vw}$ |
| | 3 | $N_{ew}\|N_e - P_o - D$ | |
| V | 1 | $P$ | |
| | 2 | $U_n^1 - P - $ Segment composed of $R_{vw} - G$ | $U_n^1 \neq G\|P$ |
| | 3 | $P - R_w - G$ | |
| | 4 | $P - EDRC2 - G$ | |
| | 5 | $P - EDRC1$ | |
| | 6 | $P - G$ | |
| | 7 | $P - R_u3\|R_u4$ | |
| VI | 1 | $U_n^1 - N_e - N_e - U_n^2 - U_n^3$ | $U_n^1 \neq R_{vw}\|G\|P\|P_o, U_n^3 \neq P_o\|N_e, U_n^2 = S\|Q\|N\|T$ |
| | 2 | $U_n^1 - N_e - N_e - N_e - U_n^2$ | $U_n^1 \notin EDRC1\|2, U_n^2 \notin EDRC1\|2$ |

| | 3 | $N_e - N_{ew} - N_{ew} - N_e$ | |
|---|---|---|---|
| | 4 | $P_o - P_o - P_o$ | |
| | 5 | $U_n^1 - U_n^2 - U_n^3 - U_n^4 - U_n^5$ | $U_n^1 \neq R_{vw}, U_n^2 = S\|Q\|N\|T, U_n^3 = S\|Q\|N\|T\|E\|D,$ $U_n^4 = S\|Q\|N\|T, U_n^5 \neq R_{vw}$ |
| | 6 | $U_n^1 - N_e - U_n^2 - U_n^3$ | $U_n^1 \neq N_e\|P_o, U_n^2 = S\|Q\|N\|T, U_n^3 = E\|D\|S\|Q\|N\|T$ |
| | 7 | $N_{ea} - W - N$ | |
| | 8 | $N_e - Y - N$ | |

Table S6 Codes for turn prediction.

**Note S9. Codes for α helix prediction**

Through analyzing protein structures in the PDB, we identified at least exist three code types that could result in formation of an α helix. The three codes and their requirements are shown in Table S5.

| | | | |
|---|---|---|---|
| I | 01 | $U_n^1 - P_o - N_e - R_{vw}\|R_w\|N_e - N_e$ | $U_n^1 \neq G, P$   $P_o \notin EDRC + G$ |
| | 02 | $U_n^1 - P_o - N_{ew} - R_{vw} - N_e^1 - N_e^2\|N_{ew} - U_n^2$ | $U_n^1\|U_n^2 \neq G, P$   $N_e^1 - N_e^2 \neq D - D$ $U_n^1 \neq N_e$   $U_n^2 \neq P_o$ |
| | 03 | $U_n^1 - P_o - N_e\|N_{ew} - R_{vw}\|C - R_{vw}\|R_w\|N_{ew} - N_e - U_n^2$ | $U_n^1\|U_n^2 \neq G, P$ |
| | 04 | $U_n^1 - P_o - N_e^1 - R_{vw}\|R_w - N_e^2\|N_{ew} - N_e^3\|N_{ew} - U_n^2$ | $U_n^1\|U_n^2 \neq G, P$   $N_e^1 - N_e^3 \neq D - D$ |
| | 05 | $U_n^1 - P_o - N_e - N_{ew} - R_{vw} - N_e\|N_{ew}$ | $U_n^1\|U_n^2 \neq G, P, N_e$   $P_o \neq H$ |
| | 06 | $U_n^1 - P_o - N_{ew} - R_{vw} - R_{vw} - P_o$ | $U_n^1 \neq P$ |
| | 07 | $U_n^1 - P_o^1 - N_e^1 - N_e^2\|N_{ew}\|R_w - R_{vw} - P_o^2$ | $U_n^1 \neq P$   $N_e^1 - N_e^2\|N_{ew} \neq D - D$ $P_o^1 - P_o^2 \neq R - R$ $P_o^1\|P_o^2 \notin EDRC + G$   $U_n^1 - N_e^1 \neq R - D$ |
| | 08 | $U_n^1 - N_{ew}^1 - P_o - R_{vw}\|M\|C - R_{vw}\|R_w\|N_e^1\|N_{ew}^2 - N_e^2\|N_{ew}^3 - U_n^2$ | $U_n^1\|U_n^2 \neq G, P$   $P_o \neq H$ $N_e^2\|N_{ew}^3 \notin EDRC$ $N_e^1\|N_{ew}^2 - N_e^2\|N_{ew}^3 \neq N_e - N_{ew}$ $U_n^1 - P_o \neq D - R$ |
| | 09 | $U_n^1 - N_e^1 - P_o - R_{vw}\|R_w - R_{vw}\|R_w\|N_e^2\|N_{ew}^1 - N_e^3\|N_{ew}^2 - U_n^2$ | $U_n^1\|U_n^2 \neq G, P$   $U_n^1 \neq P_o$   $P_o \neq H$ $N_e^2\|N_{ew}^3 \notin EDRC$ $N_e^2\|N_{ew}^1 - N_e^3\|N_{ew}^2 \neq N_e - N_{ew}$ $U_n^1 - P_o \neq D - R$   $N_e^1 - N_e^2 \neq D - D$ |
| | 10 | $U_n^1 - N_e - P_o - R_{vw}\|R_w\|N_{ew} - R_{vw}\|R_w - P_o - U_n^2$ | $U_n^1\|U_n^2 \neq G$   $U_n^1 \neq P_o$ |
| | 11 | $U_n^1 - N_{ew} - P_o^1 - R_{vw} - R_{vw}\|M\|C - P_o^2 - U_n^2$ | $U_n^1\|U_n^2 \neq G$   $U_n^1 \neq P_o$   $P_o^1 \neq H$ |
| | 12 | $U_n^1 - P_o - R_{vw} - R_{vw} - N_e - U_n^2$ | $U_n^1\|U_n^2 \neq G, P$   $P_o \notin EDRC + G$ $N_e \notin D - U_n - R$ |
| | 13 | $P_o^1 - P_o^2 - R_{vw}\|R_w - N_e^1\|N_{ew} - N_e\|N_{ew} - U_n^1$ | $U_n^1 \neq G$ $P_o^1 - P_o^2 \neq R - R \text{ except } N_e^1 = D$ |
| | 14 | $U_n^1 - P_o - R_{vw}\|R_w\|G - R_{vw}\|R_w - N_e^1 - N_e^2$ | $U_n^1 \neq G$ $N_e^1 - N_e^2 \neq D - D \text{ except } P_o = R$ |
| | 15 | $P_o - G - R_{vw} - R_{vw}\|M\|G - N_e - U_n^1$ | $U_n^1 \neq P$ |
| | 16 | $U_n^1 - N_e - P_o - R_{vw} - R_{vw} - N_e - U_n^2$ | $U_n^1\|U_n^2 \neq P, G$   $U_n^1 - P_o \neq D - R$ $U_n^2 \notin codeI \text{ for turns}$ $U_n^2 \neq P_o$ |
| | 17 | $U_n^1 - P_o - N_e - R_{vw} - R_{vw} - R_{vw} - N_e - U_n^2$ | $U_n^1\|U_n^2 \neq P, G$ |
| II | 01 | $U_n^1 - R_{vw} - R_{vw}\|M\|C - P_o\|N_e\|N_{ew} - R_{vw}\|M\|C - R_{vw} - U_n^2$ | $U_n^1\|U_n^2 \neq P$ $U_n^1 - P_o\|N_e\|N_{ew} \neq P_o - N_e\|N_{ew} \text{ or } N_e\|N_{ew} - P_o$ $U_n^2 - P_o\|N_e\|N_{ew} \neq P_o - N_e\|N_{ew} \text{ or } N_e\|N_{ew} - P_o$ $U_n^1\|U_n^2 \notin codeI \text{ for turns}$ |
| | 02 | $U_n^1 - R_{vw}^1 - P_o\|N_e\|N_{ew}\|Y\|W - P_o\|N_e\|N_{ew}\|Y\|W - R_{vw}\|M\|C - R_{vw}\|C - U_n^2$ | $U_n^1\|U_n^2 \neq G, P$   $R_{vw}^1 \neq A$ $P_o\|N_e\|N_{ew} - P_o\|N_e\|N_{ew} \notin EDRC$ $U_n^1\|U_n^2 \notin EDRC$ |
| | 03 | $U_n^1 - R_{vw}^1 - P_o\|N_e\|N_{ew} - R_{vw}\|M - R_{vw} - R_{vw} - U_n^2$ | $U_n^1\|U_n^2 \neq P$   $U_n^1 \neq G$   $R_{vw}^1 \neq A$ $U_n^1\|U_n^2 \notin EDRC$ $U_n^1 - P_o\|N_e\|N_{ew} \neq P_o - N_e\|N_{ew} \text{ or } N_e\|N_{ew} - P_o$ $U_n^2 - P_o\|N_e\|N_{ew} \neq P_o - P_o \text{ or } N_e - N_e$ $U_n^1\|U_n^2 \notin codeI, VI \text{ for turns}$ $P_o\|N_e\|N_{ew} - U_n^2 \neq P_o - P_o \text{ or } N_e - N_e$ |
| | 04 | $U_n^1 - R_{vw} - R_{vw}\|M\|C - R_{vw}\|R_w - R_{vw}\|M\|C - R_{vw} - U_n^2$ | $U_n^1\|U_n^2 \neq G, P$ $U_n^1\|U_n^2 \notin codeI, III \text{ for turns}$ |
| III | 01 | $U_n^1 - C - U_n^2 - U_n^3 - U_n^4 - U_n^5 - C - U_n^6$ | $U_n^1\|U_n^6 \neq G, P$ |

Table S7 Codes for α helix prediction.

**Note S10. Validity and efficiency of the developed codes**

To investigate the validity and efficiency of the developed codes in predicting secondary structures, we compared our predictions of 416 proteins with those from experiments in the PDB (see Table S8). The results for eight proteins are illustrated in Fig. S3. In accordance with Tables S6 and S7, the amino acids leading to α helixes and turns are highlighted. The results show that these codes have a high the success rate for prediction of an α helix (about 80%) and a turn (about 95%) in these proteins.

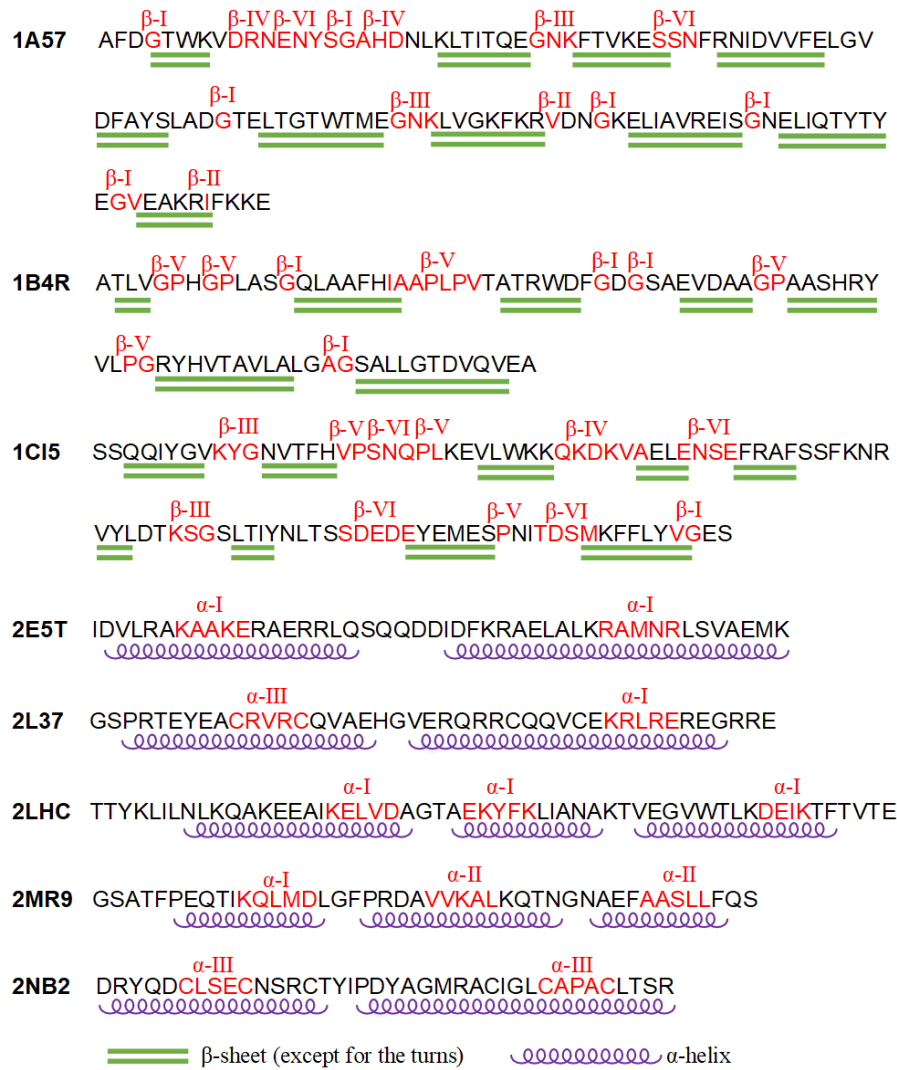

Figure. S3. Results for 8 proteins

| Proteins used for validating the folding codes of α-helix | | | | | | | | | |
|---|---|---|---|---|---|---|---|---|---|
| 2MDP | 2MH8 | 2MQJ | 2MSW | 2MX0 | 2MX2 | 2MZ0 | 2MZJ | 2N1V | 2N1W |
| 2N2Q | 2N2R | 2N2T | 2N2U | 2N3Z | 2N4C | 2N6E | 2N7J | 2N7Y | 2N9K |
| 2N12 | 2N71 | 2N76 | 2NA2 | 2NB2 | 2ND3 | 2RUE | 2RVF | 4QYW | 4ZAI |
| 4ZMD | 5AIW | 5DFG | 5JI4 | 5JN6 | 5JYU | 5KPH | 5KS5 | 5L7P | 5NCE |
| 5NPA | 5NPG | 5OMT | 5SZW | 5T1N | 5UNK | 5UP5 | 5Y6H | 5Z2S | 2KHE |
| 2KHX | 2KLZ | 2KNZ | 2KOX | 2KP7 | 2KRB | 2KRK | 2KT2 | 2KTA | 2KV8 |
| 2KXG | 2KYZ | 2KZJ | 2KZR | 2KZV | 2L0Q | 2L2M | 2L2N | 2L4B | 2L6Q |
| 2L7K | 2L9R | 2L33 | 2L37 | 2LBB | 2LE4 | 2LGI | 2LHC | 2LHE | 2LK2 |
| 2LL3 | 2LLD | 2LMZ | 2LN3 | 2LQJ | 2LQX | 2LR2 | 2LR3 | 2LR5 | 2LR8 |
| 2LRA | 2KSG | 2LT8 | 2LUQ | 2LVN | 2LXI | 2LY3 | 2LY9 | 2LYX | 2M2J |
| 2M4I | 2M05 | 2M7S | 2M66 | 2MC5 | 2MEW | 2MGV | 2MH2 | 2MTL | 2RQL |
| 2RRE | 2RRN | 2RU9 | 2W9Q | 2WNM | 2WQG | 3A5E | 3ADG | 3ADJ | 3LLB |
| 3U7T | 3ZZP | 4BWH | 4C7Q | 4C26 | 4HCS | 4N6T | 4OD6 | 1U3M | 1VCS |

| 1VMC | 1W4K | 1X4O | 1X6C | 1YG0 | 1YJR | 1ZPW | 1ZVG | 1ZXH | 2AYM |
|------|------|------|------|------|------|------|------|------|------|
| 2B7T | 2B7V | 2B68 | 2BN8 | 2C7H | 2CK4 | 2D9Y | 2DDL | 2DMU | 2DWF |
| 2E3G | 2E5T | 2EEM | 2FD9 | 2FGG | 2FHT | 2FN5 | 2GKT | 2GMG | 2GV1 |
| 2HDL | 2HJ8 | 2HJJ | 2HLU | 2IKD | 2J53 | 2JNH | 2JP6 | 2JPI | 2JQ9 |
| 2JRL | 2JSV | 2JUA | 2JUF | 2JVE | 2JVF | 2JVR | 2JWU | 2JZ5 | 2KOP |
| 2KLH | 2K2A | 2K3W | 2K4U | 2K9D | 2K49 | 2KAC | 2KAF | 2KC9 | 2KDM |
| 2KJW | 2KL1 | 2KL8 | 2KQ4 | 2NMQ | 2OED | 2OFH | 2ON8 | 2ONQ | 2PLP |
| 2ROG | 2UVS | 2V75 | 2VH7 | 2VXD | 2W4C | 2ZRR | 2ZW1 | 3CZC | 3DJN |
| 3E7U | 3G19 | 1POG | 1POU | 1PRB | 1PRU | 1PRV | 1PUZ | 1PV0 | 1Q1V |
| 1Q2N | 1R1B | 1R2A | 1R4G | 1R63 | 1RQ6 | 1RQT | 1RYK | 1SG7 | 1SGG |
| 1SKT | 1SQ8 | 1SS1 | 1SXD | 1T6O | 1TIZ | 1TNS | 1TP4 | 1TTY | 1UCP |
| 1UGO | 1UHM | 1UHS | 1USS | 1UXC | 1UXD | 1UZC | 1V63 | 1V66 | 1V92 |
| 1W3D | 1WGW | 1X2H | 1X4O | 1X4P | 1X58 | 1Z1V | 1Z96 | 1ZAC | 2AF8 |
| 2BCA | 2BCB | 2CJJ | 2CKX | 2COB | 2COS | 2CP8 | 2CP9 | 2CRA | 2CUF |
| 2DAH | 2DI0 | 2DKY | 2DNA | 2DO1 | 2E1O | 2EDU | 2EZK | 2EZL | 2FCE |
| 2GAQ | 2GDW | 2DGX | 2H80 | 2HP8 | 2J5O | 2JGV | 2JW2 | 2JWD | 2JWT |
| 2P5K | 2SPZ | | | | | | | | |
| **Proteins used for validating the folding codes of turns (including β sheet turns)** | | | | | | | | | |
| 1A57 | 1B4R | 1CI5 | 1FGP | 1G5W | 1G6P | 1GJX | 1HEJ | 1HOE | 1LPJ |
| 1LWR | 1M3A | 1M3B | 1MJC | 1MVG | 1NZ9 | 1OQK | 1PC0 | 1SA8 | 1T8V |
| 1TEN | 1TIT | 1TIU | 1TTG | 1XAK | 1YEZ | 2AVG | 2K57 | 2NCM | 2XBD |
| 3IFB | 4AIT | 1COD | 1CVO | 1F94 | 1G6M | 1G6P | 1IJC | 1JGK | 1NOR |
| 1S1N | 1TFS | 1XI7 | 1YEZ | 1ZAD | 2CRS | 2JQP | 2K57 | 2LGN | 2LXJ |
| 2MJ4 | 2O2W | 3HJD | 3LO9 | 5LUE | 5NQ4 | 1AFP | 1C01 | 1TPM | 1WR4 |
| 1ZEQ | 2JTF | 2L55 | 2LU7 | 2MW9 | 3HJD | 1AKP | 1BPV | 1C01 | 1CDB |
| 1CI5 | 1CO1 | 1DG4 | 1E0L | 1E0N | 1ED7 | 1EXH | 1G5V | 1G6E | 1G6P |
| 1G9E | 1H95 | 1HOE | 1HZK | 1I6C | 1JJX | 1K76 | 1K85 | 1KFZ | 1LWR |
| 1M3A | 1MVG | 1NZ9 | 1OH1 | 1OQK | 1OWW | 1PC0 | 1PD6 | 1POQ | 1RL1 |
| 1SHG | 1SSO | 1TEN | 1TIT | 1TIU | 1TTF | 1WIU | 1WKT | 1X32 | 1YEZ |
| 1YWJ | 1Z66 | 1ZR7 | 2AVG | 2C34 | 2CUI | 2EXD | 2F09 | 2FO8 | 2G0K |
| 2GG1 | 2JO6 | 2K3B | 2RN8 | | | | | | |

Table S8 Proteins used for validating the folding codes